\documentstyle[twocolumn]{jpsj}
\input epsf.tex

\def\runtitle{Ground-State Properties of Extended Two-Channel Kondo Model}
\def\runauthor{Hiroaki {\sc Kusunose}}

\def\hs{\mbox{\hspace{2mm}}}

\title
{
Ground-State Properties of Extended Two-Channel Kondo Model
}

\author
{
Hiroaki {\sc Kusunose}\footnote{E-mail: kusu@cmpt01.phys.tohoku.ac.jp}
}

\inst
{
Department of Physics, Tohoku University, Sendai 980-77
}

\recdate
{
October 2, 1997
}

\abst
{
Ground-state properties are examined for an extended two-channel Kondo model where the Hilbert space of the localized states is extended to include a singlet state in addition to the doublet states.
By means of zero-th order variational wavefunctions with different
symmetries, which are associated with the non-Fermi-liquid and the Fermi-liquid ground states, we demonstrate that the channel exchange coupling via the localized singlet state stabilizes the Fermi-liquid wavefunction. The ground-state phase diagrams, which are in qualitative agreement with the previous study performed by Koga and Shiba, are obtained.
The comparison to the structure of the resultant wavefunctions suggests that
a unique non-Fermi-liquid (Fermi-liquid) fixed point exists, irrespective of
the localized ground state.
}

\kword
{
quadrupolar Kondo effect, crystalline electric field, extended two-channel Kondo model, non-Fermi liquid, Fermi liquid, variational method
}

\begin{document}
\sloppy
\maketitle

In a number of heavy-fermion alloys, remarkable deviations from the Fermi-liquid behavior have been recently observed during the course of investigations of exotic heavy-fermion compounds.\cite{maple,amitsuka,lohneysen}
This non-Fermi-liquid behavior can be initiated by several microscopic origins due to, for example, the two-channel quadrupolar Kondo effect,\cite{cox} a distribution of Kondo temperatures\cite{miranda} or the critical phenomena in the vicinity of a zero-temperature quantum phase transition.\cite{hertz,millis,continentino,moriya}
In most of anomalous alloys, intersite correlations or disorder effects are not negligible to understand the non-Fermi-liquid behavior; however, a pure single-ion mechanism seems to be essential in some dilute alloys.

In particular, U$_x$Th$_{1-x}$Ru$_2$Si$_2$ $(x\le 0.07)$ shows unusual $\ln T$ dependence in the specific-heat coefficient $\gamma$, the $c$-axis magnetic susceptibility $\chi$ and the $a$-axis resistivity $\rho$.\cite{amitsuka} A couple of experimental results scaled with uranium concentration convinced us that the anomalous behaviors are due to an intrinsic single-ion effect, in particular to the magnetic two-channel Kondo effect.

On the other hand, U$_x$La$_{1-x}$Ru$_2$Si$_2$ exhibits an ordinary Kondo effect,\cite{marumoto} in which an uranium ion sits on a crystalline electric field (CEF) similar to that of the Th substituted material under tetragonal symmetry.
The difference between the two systems is the location of the first excited CEF level under the $5f^2$ configuration. The first singlet excited state is located at around 25K above the time-reversal ($\Gamma_{t5}$) ground doublet in U$_x$La$_{1-x}$Ru$_2$Si$_2$, while it is located at about $10^3$K in U$_x$Th$_{1-x}$Ru$_2$Si$_2$.

In the case of the CEF-singlet ground state, which may be realized in the hexagonal compound UPt$_3$, the low-energy physics, even for an impurity case, are still unclear; in contrast, extensive experimental investigations have been performed in order to understand exotic behaviors such as the unconventional superconductivity and extremely weak antiferromagnetism.\cite{tou}

From a theoretical point of view, several authors have investigated the influence of the CEF-singlet state on the non-Fermi-liquid fixed point of two-channel Kondo model using the numerical renormalization-group (NRG) method.\cite{koga,sakai}
They revealed that the mixing between the two channels via the CEF-singlet state plays an important role for stabilizing the Fermi-liquid fixed point over the non-Fermi-liquid one.
Although the properties of the Fermi-liquid fixed point are mainly examined, it is not yet apparent how to stabilize the non-Fermi-liquid fixed point.

In this letter, it is demonstrated that two ground states with different symmetries can be obtained on the basis of the variational treatment due to a level-crossing between them.\cite{yosida,varma,gunnasson} This approach is of great advantage for understanding how to stabilize the non-Fermi-liquid state over the Fermi-liquid state.

We begin with an extended two-channel Kondo model for tetragonal and hexagonal symmetries, in which the time-reversal doublet $|E_\pm\rangle$ and the CEF-singlet $|B\rangle$ among the $J=4$ multiplet are taken into account and only the lowest $f^1$ intermediate state has been used in the Schrieffer-Wolff transformation to derive the exchange Hamiltonian.\cite{koga} Within these restrictions, four states, which are specified by channel and spin indices, among the $j=5/2$ partial wave of conduction electrons are coupled with the internal degrees of
freedom at the impurity site.

Here it is convenient to introduce the local operators for the spin and the channel of conduction electrons defined as
\begin{eqnarray}
&&{\mib \sigma}_m = \sum_{{\mibs k}\sigma}\sum_{{\mibs k}'\sigma'}a^\dagger_{{\mibs k}m\sigma}{\mib \sigma}_{\sigma\sigma'}a_{{\mibs k}'m\sigma'},\\
&&{\mib \tau}_\sigma = \sum_{{\mibs k}m}\sum_{{\mibs k}'m'}a^\dagger_{{\mibs k}m\sigma}{\mib \sigma}_{mm'}a_{{\mibs k}'m'\sigma},
\end{eqnarray}
respectively.
Then the extended two-channel Kondo model is given in terms of both
operators by
\begin{eqnarray}
&&H=\sum_{{\mibs k}m\sigma}\epsilon_k a^\dagger_{{\mibs k}m\sigma}a_{{\mibs k}m\sigma}+\sum_{ij}^{E_+,B,E_-}\hat{M}_{ij}|i\rangle\langle j|,\\
&&\hat{M}=\frac{1}{N}\left(\begin{array}{ccc}
J(\sigma^z_1+\sigma^z_2)/2 & K\tau^-_\uparrow & J(\sigma^-_1+\sigma^-_2)\\
K\tau^+_\uparrow & N\Delta & K\tau^-_\downarrow \\
J(\sigma^+_1+\sigma^+_2) & K\tau^+_\downarrow & -J(\sigma^z_1+\sigma^z_2)/2
\end{array}\right),\nonumber
\end{eqnarray}
where $a_{{\mibs k}m\sigma}$ annihilates the ${\mib k}$-Bloch state of conduction electrons with the spin $\sigma$ ($=\uparrow,\;\downarrow$) in the channel $m$ ($=1,\;2$).
The kinetic energy $\epsilon_k$ spreads over $-D\sim +D$, $D$ being the half of a bandwidth.
The exchange couplings $J$, $K$ and $|\Delta|$, which represent the energy splitting between the doublet state $|E_\pm\rangle$ and the singlet state $|B\rangle$, are measured in units of $D$.
 The definition for each state in terms of the azimuthal component of total angular momentum, $J_z$ or $j_z$, is summarized in Table~\ref{table:1}.
\begin{table}
\caption{Definition for each state to describe the extended two-channel Kondo model. $|m\rangle$ represents the azimuthal component of total angular
momentum.}
\label{table:1}
\begin{tabular}[t]{@{\hspace{\tabcolsep}\extracolsep{\fill}}clll} \hline
State & \multicolumn{1}{c}{Tetragonal} & \multicolumn{2}{c}{Hexagonal}\\ \hline
$E_\pm$ & $\alpha|\mp3\rangle+\beta|\pm1\rangle$ & \multicolumn{2}{c}{$\alpha|\pm4\rangle+\beta|\mp2\rangle$} \\
$B$ & $(|2\rangle+|-2\rangle)/\sqrt{2}$ & \multicolumn{2}{c}{$(|3\rangle+|-3\rangle)/\sqrt{2}$} \\
$a_{{\mibs k}1\uparrow}$ & $\alpha|5/2\rangle+\beta|-3/2\rangle$ &
$|3/2\rangle$ & $|-3/2\rangle$ \\
$a_{{\mibs k}1\downarrow}$ & $|1/2\rangle$ & $|-1/2\rangle$ & $|5/2\rangle$ \\
$a_{{\mibs k}2\uparrow}$ & $|-1/2\rangle$ & $|1/2\rangle$ & $|-5/2\rangle$ \\
$a_{{\mibs k}2\downarrow}$ & $\alpha|-5/2\rangle+\beta|3/2\rangle$ &
$|-3/2\rangle$ & $|3/2\rangle$ \\ \hline
\end{tabular}
\end{table}

The second term in eq. (3) is reduced to the usual two-channel Kondo exchange interaction with the isolated singlet state $|B\rangle$ in the case of $K=0$. Nonzero $K$ term plays the role of mixing the channel degrees of freedom and the localized doublet state is simultaneously transferred to the singlet state and vice
versa.
The number of conduction electrons in each channel is no longer conserved, however, the total number of electrons,
\begin{equation}
Q=\sum_{{\mibs k}m\sigma}\left(a^\dagger_{{\mibs k}m\sigma}a_{{\mibs k}m\sigma}-\frac{1}{2}\right),
\end{equation}
is still conserved.

Since the Hamiltonian (3) is invariant under the transformation $(m\sigma E_\pm)\leftrightarrow(\bar{m}\bar{\sigma}E_\mp)$, namely, due to the time-reversal symmetry in the original representation of total angular momentum as given in Table~\ref{table:1}, we can revive a conserved quantity,\cite{koga} total spin, $J_z=\sum j_z$ by assigning fictitious spin $j_z$ to each state as
\begin{eqnarray}
&&(1\uparrow)\rightarrow(+\frac{3}{2}),\hs (2\uparrow)\rightarrow(+\frac{1}{2}),\hs (1\downarrow)\rightarrow(-\frac{1}{2}),\nonumber\\
&&(2\downarrow)\rightarrow(-\frac{3}{2}),\hs (E_\pm)\rightarrow(\pm 1),\hs (B)\rightarrow(0).
\end{eqnarray}

Now we construct possible variational wavefunctions for ground states of the extended two-channel Kondo model. A half-filled $Q=0$ state with spin polarization of conduction electrons around the impurity will be a ground state. The non-Fermi-liquid (NFL) ground state must include the overscreening state, $a^\dagger_{{\mibs k}1\downarrow}a^\dagger_{{\mibs k}'2\downarrow}|+1\rangle$ (only one of the time-reversal pair is presented), which is involved in $(Q,J_z)=(0,-1)$ state, $|j_z\rangle$ being the localized state $j_z$ and the state filled inside the
Fermi sphere to the extent $k_{\rm F}$.
While the nondegenerate Fermi-liquid (FL) state should be specified by the conserved quantities, $(0,0)$.

The variational wavefunctions can be obtained\cite{yosida} if we begin with the states in which two excited electrons above the Fermi level couple with the localized spin state and then collect higher-order states with particle-hole excitation which are connected with the next lower-order one by the exchange terms in eq. (1).
We can write a zeroth order wavefunction $|0,-1\rangle$ for NFL and $|0,0\rangle$ for FL states in consideration of the time-reversal symmetry, $J_z\leftrightarrow-J_z$, as
\begin{eqnarray}
&&|0,-1\rangle = \nonumber\\
&&\mbox{\hspace{3mm}}\sum_{1,2}^{\ge k_{\rm F}}\biggl[
\alpha^{(1)}_{12}|{\mib k}_1 1\uparrow,{\mib k}_2 2\downarrow;-1\rangle
+\alpha^{(2)}_{12}|{\mib k}_1 2\uparrow,{\mib k}_2 1\downarrow;-1\rangle\nonumber\\
&&\mbox{\hspace{5mm}}+\alpha^{(3)}_{12}|{\mib k}_12\downarrow,{\mib k}_21\downarrow;+1\rangle
+\alpha^{(4)}_{12}|{\mib k}_12\downarrow,{\mib k}_22\uparrow;0\rangle\biggr],\\
&&|0,0\rangle = \nonumber\\
&&\mbox{\hspace{3mm}}\sum_{1,2}^{\ge k_{\rm F}}\biggl[
\frac{1}{\sqrt{2}}\beta^{(1)}_{12}\biggl(
|{\mib k}_11\uparrow,{\mib k}_21\downarrow;-1\rangle
-|{\mib k}_12\downarrow,{\mib k}_22\uparrow;+1\rangle\biggr)\nonumber\\
&&\mbox{\hspace{5mm}}+\frac{1}{2}\beta^{(2)}_{12}\biggl(
|{\mib k}_11\downarrow,{\mib k}_21\downarrow;+1\rangle
-|{\mib k}_12\uparrow,{\mib k}_22\uparrow;-1\rangle\biggr)\nonumber\\
&&\mbox{\hspace{5mm}}+\beta^{(3)}_{12}|{\mib k}_12\downarrow,{\mib k}_21\uparrow;0\rangle
\biggr],
\end{eqnarray}
where we define $|{\mib k}_1m\sigma,{\mib k}_2m'\sigma';j_z\rangle=a^\dagger_{{\mibs k}_1m\sigma}a^\dagger_{{\mibs k}_2m'\sigma'}|j_z\rangle$ and ${\mib k}_n$ is simply expressed as $n$ in the variational coefficients, $\alpha^{(i)}$ and $\beta^{(i)}$.
The Pauli principle requires that $\beta^{(2)}_{12}=-\beta^{(2)}_{21}$ and $\beta^{(3)}_{12}=\beta^{(3)}_{21}$ is due to time-reversal symmetry.
It is noted that $|{\mib k}_11\downarrow,{\mib k}_21\downarrow;0\rangle$ ($|{\mib k}_11\downarrow,{\mib k}_22\uparrow;0\rangle$) does not connect with the above wavefunction $|0,-1\rangle$ ($|0,0\rangle$) despite the fact that its conserved quantities are the same.

If we ignore the $\mib k$ dependence of the variational wavefunctions, they are
similar to the wavefunctions which are obtained by the diagonalization of zeroth-site Hamiltonian in the NRG method.\cite{wilson}
The successive renormalization procedure can improve these localized
wavefunctions to involve the low-energy states of conduction electrons.
In our zeroth variational wavefunctions, such a strict improvement
is replaced only to take into account the $\mib k$ dependence of conduction
electrons.
\begin{figure}
\vspace{3mm}
\begin{center}
\epsfxsize=8cm \epsfbox{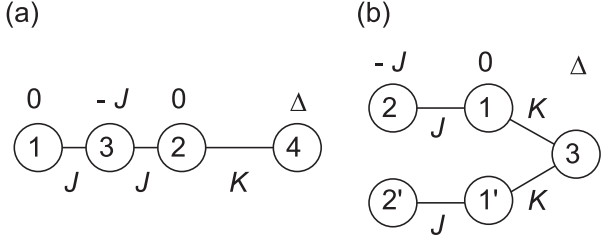}
\end{center}
\caption{The relations among each component in the variational wavefunction (a) $|0,\pm 1\rangle$, (b) $|0,0\rangle$. The circle with the number $i$ stands for the $i$-th component of the variational function whose energy is indicated above the circle (the kinetic energy $\epsilon_1+\epsilon_2$ is omitted).}
\label{fig:3}
\end{figure}

The candidates in the variational functions are connected with each other by exchange interactions. The relations among them are schematically shown in Fig.~\ref{fig:3}. The circle with the number $i$ stands for the $i$-th component of the variational function whose energy is indicated above the circle (the kinetic energy $\epsilon_1+\epsilon_2$ is omitted).
We can easily expect that $J$ stabilizes the state $(0,\pm 1)$, while the state $(0,0)$ can be stabilized due to the exchange $K$ because the latter state has two mixing paths with respect to $K$, while in the former state, two mixing paths are available for $J$ instead of $K$.
Since the relation of the number of mixing paths remains the same when considering higher order wavefunctions, we expect that the zeroth order wavefunctions contain essential features to clarify level-crossing of ground states.

The Schr\"odinger equation $[H-E(Q,J_z)]\;|Q,J_z\rangle=0$ provides us with the following set of equations to determine $\alpha^{(i)}_{12}$ and $\beta^{(i)}_{12}$ and $E$:
\begin{eqnarray}
&&
\left\{\begin{array}{l}
\xi_{12}\alpha^{(1)}_{12}-J\left(A^{(1)}_{2-1}+\bar{A}^{(3)}_2\right)=0,\\
\xi_{12}\alpha^{(2)}_{12}-J\left(A^{(2)}_{2-1}-A^{(3)}_2\right)-2KA^{(4)}_1=0,\\
\xi_{12}\alpha^{(3)}_{12}-J\left(A^{(3)}_{2+1}+A^{(1)}_1-A^{(2)}_2\right)=0,\\
\left(\xi_{12}+\Delta\right)\alpha^{(4)}_{12}-2K\bar{A}^{(2)}_2=0,
\end{array}\right.\\
&&
\left\{\begin{array}{l}
\xi_{12}\beta^{(1)}_{12}-J\left(B^{(1)}_{2-1}-\sqrt{2}B^{(2)}_2\right)
-\sqrt{8}KB^{(3)}_1=0,\\
\xi_{12}\beta^{(2)}_{12}-J\left(B^{(2)}_{2+1}+\frac{1}{\sqrt{2}}\bigl(B^{(1)}_1-B^{(1)}_2\bigl)\right)=0,\\
\left(\xi_{12}+\Delta\right)\beta^{(3)}_{12}-\sqrt{2}K\left(\bar{B}^{(1)}_1+\bar{B}^{(1)}_2\right)=0,
\end{array}\right.
\end{eqnarray}
where $\xi_{12}=\epsilon_1+\epsilon_2-E$, and for $(C,\gamma)=(A,\alpha)$ or $(B,\beta)$ $C^{(i)}_m=(1/N)\sum_3\gamma^{(i)}_{3m}$, $\bar{C}^{(i)}_m=(1/N)\sum_3\gamma^{(i)}_{m3}$, $C^{(i)}_{m\pm n}=(C^{(i)}_m\pm \bar{C}^{(i)}_n)/2$.

First, we consider the case $J,\;K\gg D\equiv 1$, which is related to the local
problem in the NRG method as mentioned above.
We can assume $\gamma^{(i)}_{12}=\gamma^i$ (independent of ${\mib k}$); then eqs. (8) and (9) are reduced to
\begin{eqnarray}
&&\left(\begin{array}{cccc}
\zeta & 0 & -J & 0\\
0 & \zeta & J & -2K\\
-J & J & \zeta-J & 0\\
0 & -2K & 0 & \zeta+\Delta
\end{array}\right)
\left(\begin{array}{c}
\alpha^1\\ \alpha^2\\ \alpha^3\\ \alpha^4
\end{array}\right)=0,\;\;\;\;\\
&&\left(\begin{array}{cc}
\zeta & -2\sqrt{2}K\\
-2\sqrt{2}K & \zeta+\Delta
\end{array}\right)
\left(\begin{array}{c}
\beta^1\\ \beta^3
\end{array}\right)=0,
\end{eqnarray}
respectively, where $(\zeta+E)/2=\langle\epsilon_k\rangle=\sum_{\mibs k}\epsilon_k/N$ and the component of $\beta^2$ vanishes because of the antisymmetric condition.

From these equations, we can obtain ground-state energies $\bar{E}(Q,J_z)=E-2\langle\epsilon_k\rangle$ as
\begin{subeqnarray}
&&\bar{E}(0,\pm 1)=\min(-2J,\;\Delta),\\
&&\bar{E}(0,0) = \min(0,\;\Delta)\mbox{\hspace{3cm}}
\end{subeqnarray}
for $J\gg K$ and
\begin{subeqnarray}
\bar{E}(0,\pm 1)=\min\left[\frac{1}{2}\left(\Delta-\sqrt{\Delta^2+16K^2}\right),\;-J\right],\\
\bar{E}(0,0) = \frac{1}{2}\left(\Delta-\sqrt{\Delta^2+32K^2}\right)\mbox{\hspace{2.25cm}}
\end{subeqnarray}
for $J\ll K$. Thus, the ground state $(0,\pm 1)$ can be realized for the case $J\gg K,|\Delta|$ or $\Delta\gg J,K$, while $(0,0)$ for $K\gg J,|\Delta|$ or $-\Delta\gg J,K$.

\begin{figure}
\vspace{3mm}
\begin{center}
\epsfxsize=8cm \epsfbox{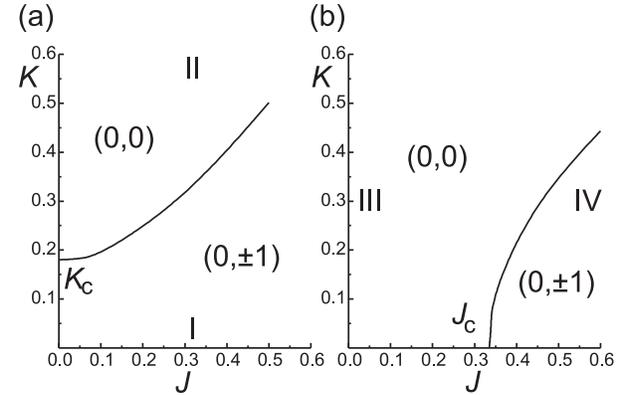}
\end{center}
\caption{The ground-state diagrams for exchange couplings $J$ and $K$: (a) $\Delta=0.2$ (b) $\Delta=-0.2$. The nature of the typical ground states can be seen in regions I--IV.}
\label{fig:1}
\end{figure}

Next, we investigate the ground state properties for various $J$ and $K$ numerically.
The variational parameters are adjusted suitably with advantage both in the exchange energies $J$, $K$ and the kinetic energy $\epsilon_k$.
The phase diagrams of the ground state are shown in Fig.~\ref{fig:1}(a) for $\Delta=0.2$ (time-reversal-doublet ground state) and 2(b) for $\Delta=-0.2$ (CEF-singlet ground state).
The boundary lines are determined by $E(0,0)=E(0,\pm 1)$ and the singlet state is stable for $E(0,0)<E(0,\pm 1)$, while the doublet state is stable for
$E(0,\pm 1)<E(0,0)$.
These results are in qualitative agreement with the previous study using the NRG calculation.\cite{koga}
The critical strength of the exchange coupling $K_{\rm c}$ for (a) or $J_{\rm c}$ for (b) is roughly determined by $|\Delta|$.
It is noted that the boundary for negative $\Delta$ is shifted to the right more than that expected as $K\sim J$ for large $J$, $K$. Typical examples for the state $(0,\pm 1)$ are given in the regions I and IV and for the state $(0,0)$ in the regions II and III in Fig.~\ref{fig:1}.

\begin{figure}
\vspace{3mm}
\begin{center}
\epsfxsize=8cm \epsfbox{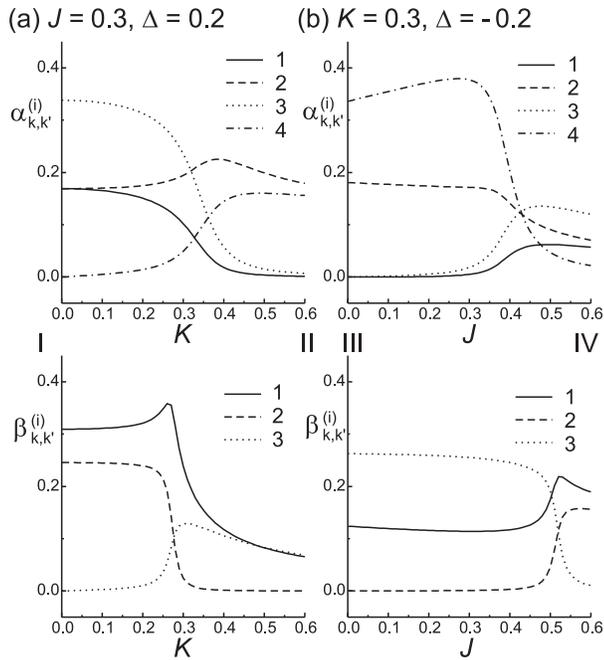}
\end{center}
\caption{Changes of variational parameters at $k=k'=k_{\rm F}$ ($k=k_{\rm F}$, $k'=k_{\rm F}+\delta$ for $\beta^{(2)}$) across the phase boundary: (a) for $J=0.3$ and $\Delta=0.2$ and (b) for $K=0.3$ and $\Delta=-0.2$.}
\label{fig:2}
\end{figure}

The changes of variational parameters at $k=k'=k_{\rm F}$ ($k=k_{\rm F}$, $k'=k_{\rm F}+\delta$ for $\beta^{(2)}$) across the phase boundary are shown in Fig.~\ref{fig:2}(a) for $J=0.3$ and $\Delta=0.2$ (I $\rightarrow$ II) and 3(b) for $K=0.3$ and $\Delta=-0.2$ (III $\rightarrow$ IV).

For typical $(0,\pm 1)$ ground state in region I, $\alpha^{(1)}\sim\alpha^{(3)}$ are dominant.
The $\alpha^{(3)}$ term associated with the overscreening state becomes most dominant as expected in the conventional two-channel Kondo model.
This situation is also realized in region IV, although the localized singlet state is stable (in this case, $\alpha^{(4)}$ term is expected to be dominant),
because the exchange energy $K$ surpasses the loss of the localized energy $|\Delta|$.
Since the distribution of intensity spreads over the higher kinetic-energy states due to the strong coupling constants, each intensity of variational parameters at the Fermi level in the latter example is smaller than that of the former example.
On the other hand, the $\beta^{(1)}$ and the $\beta^{(3)}$ terms are dominant for $(0,0)$ ground state in regions II and III, which include the decoupled CEF-singlet state as a limit $\beta^{(1)}\rightarrow 0$.\cite{koga}
This result implies that a unique non-Fermi-liquid (Fermi-liquid) fixed point
exists, irrespective of the CEF ground state.

The structure of the wavefunction changes drastically but continuously across
the phase boundary in order to make full use of the larger exchange coupling.
As a direct result of this change, one stable ground state is replaced by the
other one. The low-lying excitations between both sides of the phase boundary might be completely different because the nature of the ground state changes
abruptly at the boundary.
This situation is essentially analogous to that of an anisotropic spin $S=1$ in the magnetic field, which is described by the Hamiltonian, $H=DS_z^2-S_zB$. In this example, the ground state $|0\rangle$ for $B<D$ is replaced by $|+1\rangle$ for $B>D$.
However, our variational state has a corresponding many-body resonance in
contrast to the purely one-body state $|S_z\rangle$.

In summary, we have investigated the ground-state properties of the extended two-channel Kondo model on the basis of the variational treatment.
We construct the ground-state wavefunctions with different symmetries, which are associated with the non-Fermi-liquid and the Fermi-liquid ground states.
Within the 0th order wavefunctions, in which the particle-hole excitations are ignored, the non-Fermi-liquid ground state is realized due to the spin exchange $J$, while the channel exchange $K$ stabilizes the Fermi-liquid state both for the CEF time-reversal-doublet and the CEF-singlet ground states.
This quantum phase transition comes from the competition between the different
types of ground states with corresponding many-body resonances.
The comparison to the structure of the resultant wavefunctions suggests that
a unique non-Fermi-liquid (Fermi-liquid) fixed point exists,
irrespective of the CEF ground state.

The author would like to thank Prof. Y. Kuramoto for the useful discussions and critical reading of this manuscript.
He has benefited greatly from the stimulating correspondence with Dr. M. Koga.
He is also indebted to Prof. O. Sakai and Dr. R. Takayama for the valuable suggestions on numerical calculations.

\end{document}